\pdfoutput=1

\documentclass[journal]{IEEEtran}
\usepackage{cite}

\ifCLASSINFOpdf
\else
\fi
\usepackage{amsmath}

\usepackage{algorithm}
\usepackage{algpseudocode}
\usepackage{graphicx}
\usepackage{amsfonts}
\usepackage{booktabs}
\usepackage{subcaption}
\usepackage{float}

\begin{document}
%
\title{Reliable IoT Storage: Minimizing Bandwidth Use in Storage Without Newcomer Nodes}
%
%
%


\author{Xiaobo Zhao,
        Daniel E. Lucani,
        Xiaohong Shen,
        and~Haiyan Wang\vspace{-0.5cm}


\thanks{X. Zhao is a visiting student at Aarhus University, supported by the Chinese Scholarship Council (CSC). This research is supported partly by National Natural Science Foundation of China (61671386), partly by the Cisco University Research Program Fund (Project CG \#593761), Gift \#2015-146035 (3696) and partly by the AUFF Starting Grant Project AUFF-2017-FLS-7-1.}
\thanks{X. Zhao, X. Shen and H. Wang are with Key Laboratory of Ocean Acoustics and Sensing, Northwestern Polytechnical University, Ministry of Industry and Information Technology and School of Marine Science and Technology, Northwestern Polytechnical University, Xi’an, China. (e-mail: xbzhao@mail.nwpu.edu.cn, \{xhshen, hywang\}@nwpu.edu.cn)}
\thanks{X. Zhao and D. E. Lucani are with the Department of Engineering, Aarhus University, Aarhus, Denmark. (e-mail: xiaobo.zhao@eng.au.dk, daniel.lucani@eng.au.dk)}}
\maketitle

\begin{abstract}
This letter characterizes the optimal policies for bandwidth use and storage for the problem of distributed storage in Internet of Things (IoT) scenarios, where lost nodes cannot be replaced by new nodes as is typically assumed in Data Center and Cloud scenarios. We develop an information flow model that captures the overall process of data transmission between IoT devices, from the initial preparation stage (generating redundancy from the original data) to the different repair stages with fewer and fewer devices. Our numerical results show that in a system with 10 nodes, the proposed optimal scheme can save as much as 10.3~\% of bandwidth use, and as much as 44~\% storage use with respect to the closest suboptimal approach.
\end{abstract}

\begin{IEEEkeywords}
Network coding, distributed storage, Internet of Things, information flow
\end{IEEEkeywords}

%
\IEEEpeerreviewmaketitle

\vspace{-0.15cm}
\section{Introduction}
\IEEEPARstart{T}{he} Internet of Things (IoT) has gained increasing attention in various fields due to its ability to enable new devices to connect to the Internet~\cite{atzori2010internet}. Many IoT-based applications, e.g., environmental monitoring~\cite{miorandi2012internet}, data collection~\cite{orsino2016energy}, and industrial automation~\cite{wang2014iot}, have a common challenge, namely, how to maintain data reliably when using vulnerable, resource-limited devices. Device failures are caused by a variety of different factors, including limited energy, hardware failures, and software errors~\cite{qiu2016greedy}. IoT systems that are deployed in harsh or unaccessible environments~\cite{lazarescu2013design} face unique challenges to replace damaged devices or they will be unable to do so. Thus, it is critical to develop a theoretical framework and policies to maintain data reliably in such systems without expecting replacement devices. The goal of this paper is to provide such framework and use it to minimize communication and energy consumption costs for the IoT devices as well as reducing the data storage requirement of individual device.

Network coding~\cite{ahlswede2000network} for distributed storage has been shown to provide efficient solutions to maintain data reliably for long periods of time~\cite{dimakis2010network, guerrero2016network, zhao2017reliable}. Early work introducing network coding to Distributed Storage Systems (DSS) identified the optimal tradeoff between storage and repair cost, i.e., the network traffic that creates packets of the newcomer node~\cite{dimakis2010network}. More recently, the problem of reliability without newcomer nodes has been considered in~\cite{guerrero2016network} and \cite{zhao2017reliable}, assuming that a lead node is required for coordination purposes. The former studies how to transition optimally between extremal operation points of the optimal tradeoff curve in~\cite{dimakis2010network} without newcomers. The latter revealed that the cost for introducing redundancy at the beginning of the storage process has a critical role in the overall cost to the system. Additionally, the timing and quantity of redundancy introduced as multiple nodes suffer failures has an effect on reducing the repair cost of failures. An optimal solution for the problem described in~\cite{zhao2017reliable} is missing.

We investigate the fundamental problem of minimizing bandwidth and storage costs for the data protection with no newcomers, but without the need for a lead/coordinating node in the process.
We model the problem based on an information flow graph and formulate an optimization problem to identify the optimal process for the repair of data. We show that the problem of minimizing traffic can be formulated as a Linear Programming (LP) optimization problem. The solution of this LP problem provides an optimal strategy in terms of when and how much redundancy/traffic should be generated by the system coming from each node. Our analysis method and solution is not limited to IoT applications. By comparing to suboptimal protection strategies inspired by the work in~\cite{guerrero2016network,zhao2017reliable}, we show that significant gains can be achieved by solving the optimization problem both in bandwidth reduction, but also by reducing storage requirement per device.

\vspace{-0.1cm}
\section{System Model}

We consider an $n$-node network has $M$ bits of initial data, where each node has generated $M / n$ bits.
Nodes can fail (leave without sending data to others nodes) one at a time. We assume that the data will be recovered by a data collector at the end of the process, even if only $k$ $(1 \leq k < n)$ nodes survive.
For simplicity, the network is considered as a symmetric one where each node has the same probability to be unavailable and any $k$ nodes are sufficient to recover the original data. In the sense of network management, the costs are negligible compared with those of repairs, and network protocols have been well studied in the literature. We assume that the system operates well without considering the protocols in this letter.

The system encodes the data and distributes coded data from each device to the others to achieve a certain redundancy across the $n$ nodes.
This first stage is called the \textit{preparation phase}.
The amount of transmitted data from all nodes is called the \textit{preparation bandwidth}.
When a node fails, a repair is carried out, in which any two nodes communicate with each other to recover the lost redundancy and guarantee that subsequent node failures will not cause permanent loss of data. The amount of transmitted data from all nodes in this stage is called the \textit{repair bandwidth}.

We define the summation of preparation and repair bandwidth as the \textit{protection bandwidth}. Since the preparation phase can be regarded as a special repair phase if we consider that there is a loss of a virtual node and $n$ nodes remain after that loss, we will not distinguish between preparation bandwidth and repair bandwidth in our analysis, contrary to what was done in~\cite{zhao2017reliable}.
\begin{figure}[ht]
  \centering
  \includegraphics[width = 3.2in, height = 3cm]{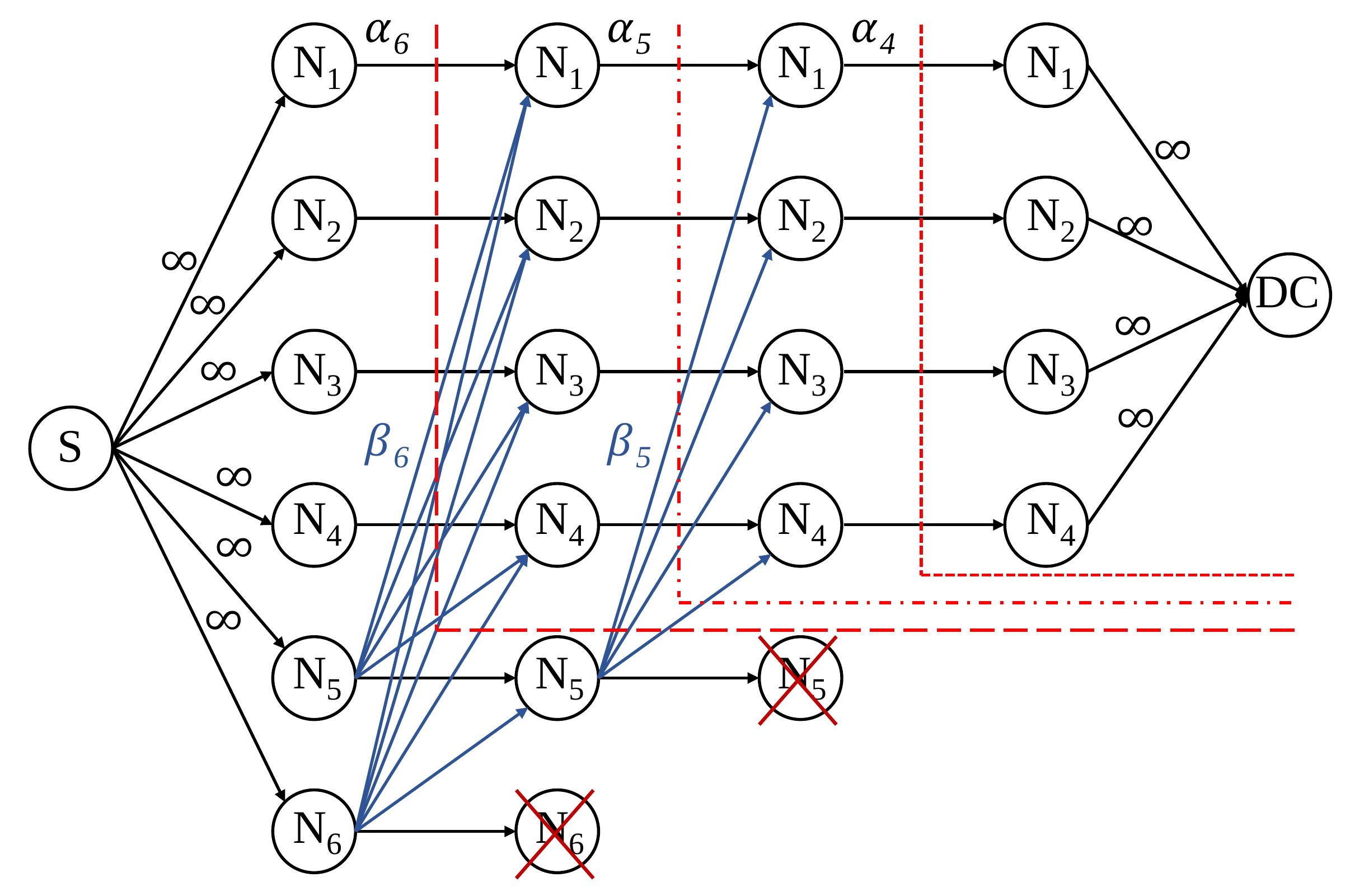}\\
  \caption{An illustration of the protection problem corresponding to $n = 6$ and $k = 4$ based on an information flow graph.}  \label{Fig1}
  \vspace{-0.5cm}
\end{figure}
\vspace{-0.1cm}

\vspace{-0.15cm}
\subsection{Performance Metrics}
Our key performance metrics are the protection bandwidth and the storage cost, as the former is mapped to the network and energy costs and the latter to a cost in the design of the IoT devices, i.e., how much storage capacity is needed. Considering a protection problem with parameters $n$ and $k$, we denote the storage in each of $i$ existing nodes by $\alpha_i$, and the storage overhead by $\sigma(n, k) = \sum_{i = k}^{n} i \alpha_i$ ($k \leq i \leq n$).

Concerning network traffic of the problem, we denote the amount of data sent from each node in a repair stage as $\beta_s$, where $s$ nodes are involved $(k + 1 \leq s \leq n)$. The repair bandwidth consists of entire $\beta_s$, denoted as $\gamma_s = s (s - 1) \beta_s$. The protection bandwidth is defined as the summation of all $\gamma$ as $\delta(n, k) = \sum_{s = k + 1}^{n} \gamma_s$. The reason to use the protection bandwidth as key metric is that the problem of repair without newcomers is different from~\cite{dimakis2010network} in that previous losses and actions to repair from those losses have an effect on subsequent repairs, i.e., the repair bandwidth depends on how many nodes remain in a repair stage, but also on previous repairs. For example, the repair bandwidth of one or more repair stages can become zero if in a previous stage the storage of each device is increased sufficiently. An extreme and wasteful case occurs when the preparation phase generates full copies of the data in all nodes, thus making any data exchange after losses unnecessary.
\vspace{-0.1cm}

\vspace{-0.2cm}
\subsection{Information Flow Graph}
In~\cite{dimakis2010network}, the original data can be reconstructed provided that the minimum capacity of the cuts separating source and data collector is equal or greater than its size. Although the formulation is not suitable in our scenario, the insight of mapping the repair problem to a multicast problem on an information flow graph is still useful. However, it requires a non-trivial adaptation to incorporate the various repair stages and a more complex set of min-cut constraints. We utilize an information flow graph to map the protection problem to a multicasting problem. Fig.~\ref{Fig1} is an example corresponding to $n = 6$, $k = 4$ to illustrate the mapping.

The information flow graph is a directed acyclic graph, which is composed of three categories of nodes: a source $\mathrm{S}$, ordinary nodes $\mathrm{N}$'s, and a data collector $\mathrm{DC}$. In terms of capacity, $\mathrm{S}$ communicates original information to each ordinary node with a directed edge of infinite capacity, and $\mathrm{DC}$ is linked to every remaining node by a directed edge with infinite capacity. The directed edges between ordinary nodes are of two patterns, and the distinction lies in whether the destination is exactly the transmitter. In a repair stage, the capacity of directed edges between the same node $(\mathrm{N_u} \rightarrow \mathrm{N_u})$ is the storage $\alpha$, and that between different nodes $(\mathrm{N_u} \rightarrow \mathrm{N_w})$ is the transmission size $\beta$. Note that notations $\mathrm{N_u}$ and $\mathrm{N_w}$ are employed to distinguish the two edge patterns, and they do not refer to specific nodes.

In our particular problem, if there is more than one failure $(n - k \geq 2)$, the min-cut of $\mathrm{S}$-$\mathrm{DC}$ cuts in each stage should be equal or greater than $M$ to guarantee the reconstruction of original data. Nevertheless, only the original data size in each node $\alpha_n$ is known, and the values of $\alpha_k, \ldots, \alpha_{n - 1}, \beta_{k + 1}, \ldots, \beta_n$ will vary with different strategies being employed. Without the capacity of edges, the capacity of a cut cannot be obtained, and then the minimum cut cannot be determined. Given this, we consider to let all $\mathrm{S}$-$\mathrm{DC}$ cuts (with finite capacity) in each stage satisfy the data reconstruction condition, then every min-cut will satisfy the condition without a doubt. The detail will be provided in next section.

Unlike the more general multicast problem considering all losses at once, we choose a sequence of events for them, which provides a better comprehension of the metric relationship for two successive repair stages.
Moreover, due to the problem's symmetry, we can focus on one path, i.e., one set of remaining nodes, with the understanding that each of the other paths will contribute to the traffic $\beta$.
As the example in Fig.~\ref{Fig1}, we focus only on the beneficial transmissions, which carry information that would be lost otherwise when losing nodes $\mathrm{N_6}$ and $\mathrm{N_5}$. The transmissions sent from other nodes will be counted in while calculating the protection bandwidth $\delta$.

\vspace{-0.2cm}
\section{Protection Bandwidth Analysis}
This section provides a systematic method to find all cuts in the information flow graph of a multicast problem, which is mapped by a protection problem with arbitrary starting point $n$ and finishing point $k$. We also establish an optimization problem to minimize the protection bandwidth for $n$ and $k$.

\vspace{-0.2cm}
\subsection{Example}
As shown in Fig.~\ref{Fig1}, by two repair stages, $\mathrm{DC}$ can reconstruct the original data of a six-node system after two nodes fail. We denote the set of all $\mathrm{S}$-$\mathrm{DC}$ cuts with five nodes remaining as $\mathrm{(S,DC)}_{(6, 5)}$, and that with four nodes remaining as $\mathrm{(S,DC)}_{(6, 4)}$. Since we focus only on $\mathrm{S}$-$\mathrm{DC}$ cuts in this letter, we rewrite $\mathrm{(S,DC)}_{(6, 5)}$ and $\mathrm{(S,DC)}_{(6, 4)}$ as $(6, 5)$ and $(6, 4)$ without ambiguity. Due to space limitations, only three typical cuts of $(6, 4)$ are shown in Fig.~\ref{Fig1}, but all additional cuts are accounted for in the analysis. And the capacity set of $(6, 5)$ is

\makeatletter
    \def\tagform@#1{\maketag@@@{\normalsize(#1)\@@italiccorr}}
\makeatother
\footnotesize
\begin{equation}
c(6, 5) = 
\left\{
\begin{aligned}
  & 5 \alpha_6 + (6 - 5) \cdot 5 \beta_6 + 0 \alpha_5\\
  & 4 \alpha_6 + (6 - 5) \cdot 4 \beta_6 + 1 \alpha_5\\
  & 3 \alpha_6 + (6 - 5) \cdot 3 \beta_6 + 2 \alpha_5\\
  & 2 \alpha_6 + (6 - 5) \cdot 2 \beta_6 + 3 \alpha_5\\
  & 1 \alpha_6 + (6 - 5) \cdot 1 \beta_6 + 4 \alpha_5\\
  & 0 \alpha_6 + (6 - 5) \cdot 0 \beta_6 + 5 \alpha_5 ~.
\label{sixfour}
\end{aligned}
\right.
\end{equation}
\normalsize
\begin{figure}[ht]
    \centering
    \begin{subfigure}{0.4\textwidth}
        \includegraphics[width=\textwidth, height = 3.2cm]{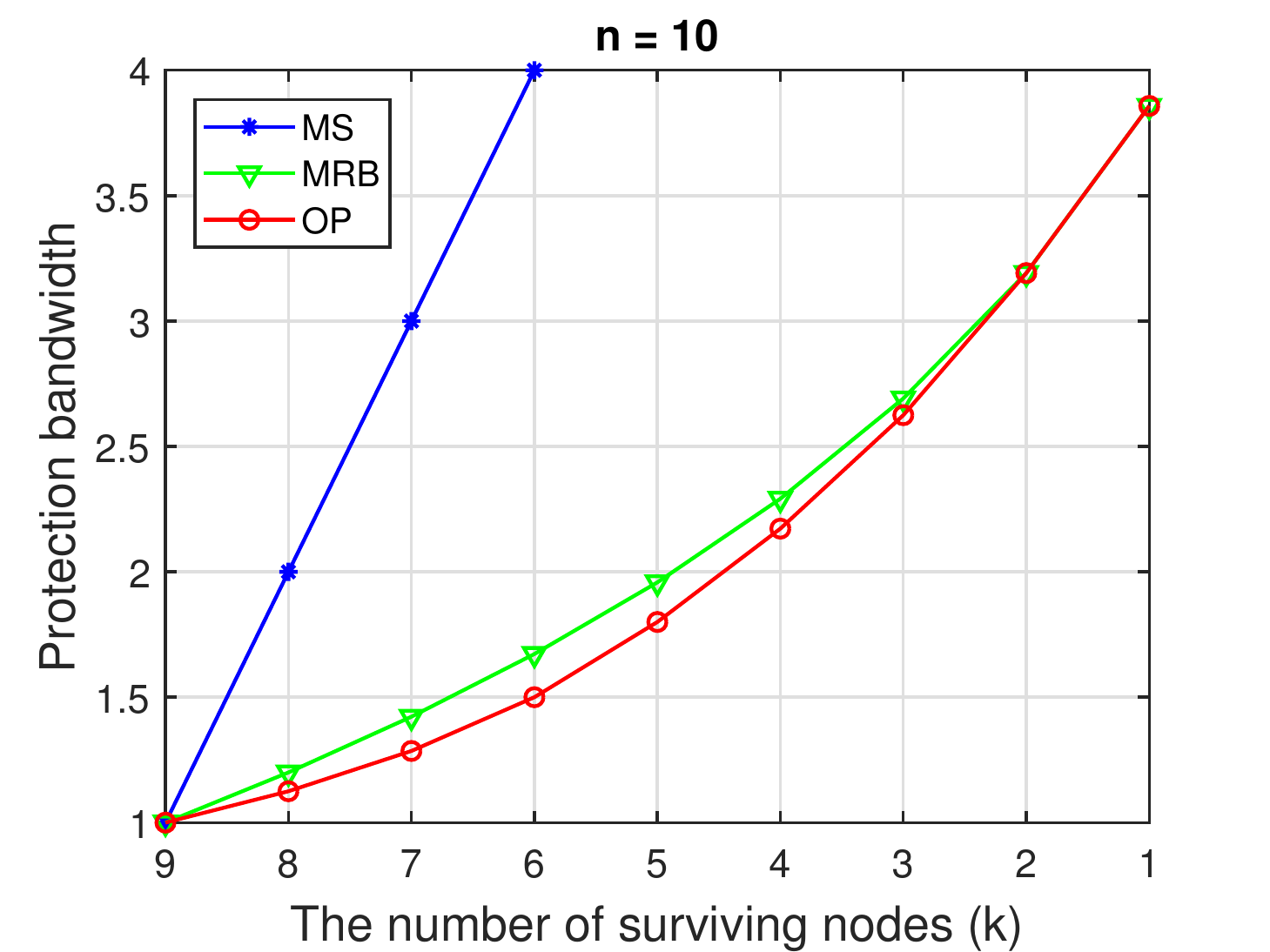}
        \caption{Protection bandwidth of MS, MRB and OP for different numbers of surviving nodes $k$, and a fixed numbers of original nodes $n = 10$.}
        \label{fig2a}
    \end{subfigure}
    \begin{subfigure}{0.4\textwidth}
        \includegraphics[width=\textwidth, height = 3.2cm]{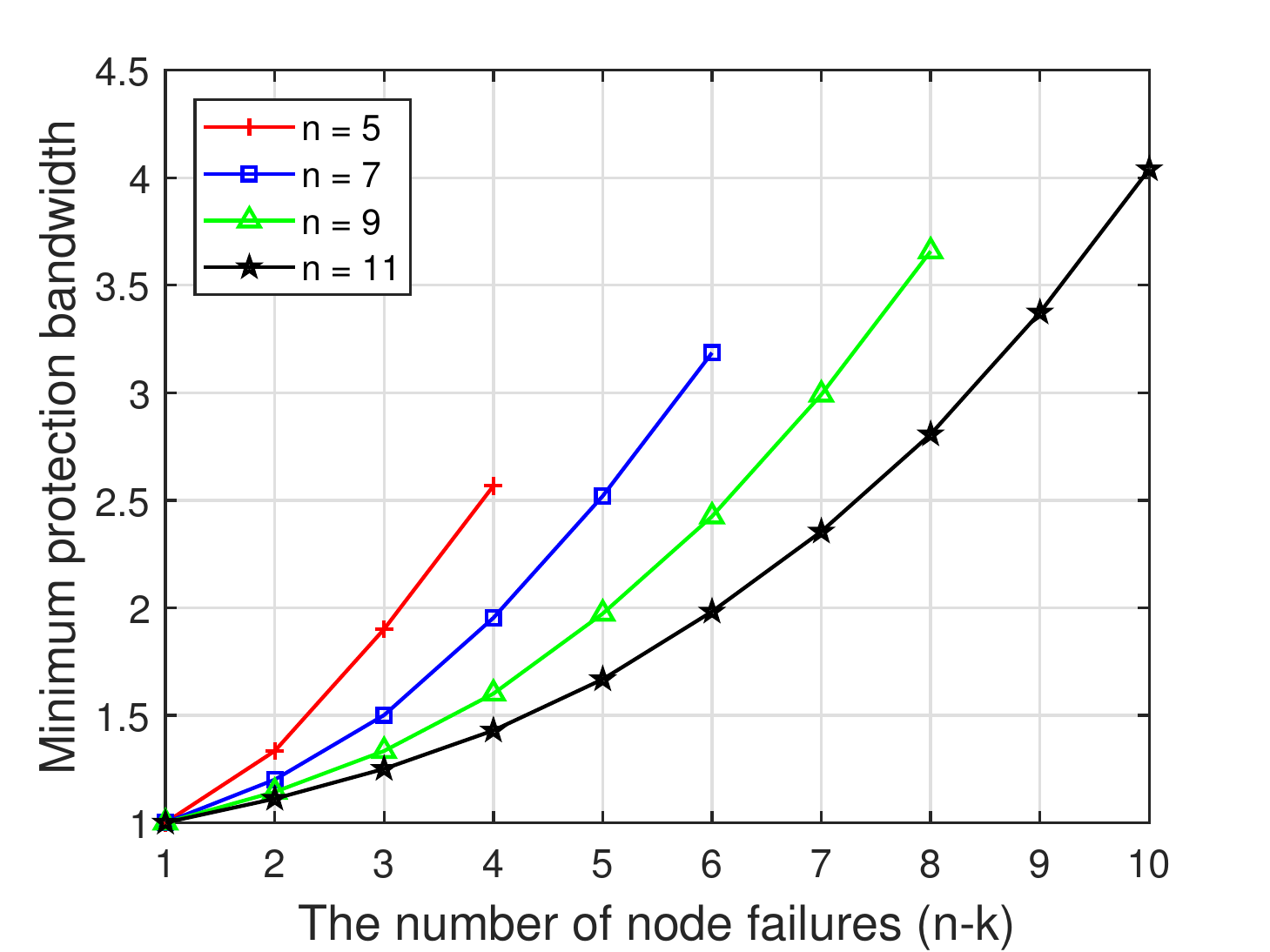}
        \caption{Minimum protection bandwidth of OP strategy for different numbers of original nodes. The number of node failures is got by $n - k$.}
        \label{fig2b}
    \end{subfigure}
    \caption{Normalized protection bandwidth of different strategies}\label{Fig2}
    \vspace{-0.6cm}
\end{figure}

Similarly, we have the capacity set of all $\mathrm{S}$-$\mathrm{DC}$ cuts as four nodes remain $c(6, 4)$. Based on the analysis before, $c(6, 5) \geq M$ and $c(6, 4) \geq M$ are necessary to reconstruct the original data, i.e., they are the constraints of cut capacity of a protection problem with $n = 6$ and $k = 4$.
Combining with the constraints of $\alpha$'s and $\beta$'s, we establish an optimization problem to minimize the protection bandwidth $\delta(6, 4)$ as

\footnotesize
\begin{equation}
\begin{aligned}
\text{min} \quad &\delta(6, 4) = 6(6 - 1) \beta_6 + 5(5 - 1) \beta_5 \\
&\text{s.t.}
\begin{cases}
c(6, 5) \geq M \\
c(6, 4) \geq M \\
\alpha_{6} + 5 \beta_{6} \geq \alpha_5 \\
\alpha_{5} + 4 \beta_{5} \geq \alpha_4 \\
0 \leq \beta_{6} \leq \alpha_{6} \\
0 \leq \beta_{5} \leq \alpha_{5} \\
\alpha_6 = \frac{M}{6} ~,
\end{cases}
\end{aligned}
\label{op_64}
\end{equation}
\normalsize
where the constraints apart from cut capacity are related to storage and repair bandwidth. In this example, $\alpha_{m} \leq \alpha_{m + 1} + m \beta_{m + 1}$ $(m = 4, 5)$ and $0 \leq \beta_{s} \leq \alpha_{s}$ $(s = 5, 6)$ limit their ranges, respectively. The principle is that the storage in a node cannot exceed the sum of the previous data already stored and the total data received from the repair stage and that each transmission cannot be negative or exceed the amount of stored data in the node. Moreover, $\alpha_6$ is known as $M / 6$.

Both of the objective function and constraints in~\eqref{op_64} are linear, thus, by solving this LP problem, we can determine the minimum protection bandwidth $\delta(6, 4)_{min}$, and the optimal protection strategy $(\beta_5, \beta_6)$ that achieves it. Next, we will formulate the problem to acquire $\delta(n, k)_{min}$.
\vspace{-0.15cm}

\vspace{-0.2cm}
\subsection{Minimum Protection Bandwidth}

For our information flow graph with arbitrary $n$ and $k$, the set of all $\mathrm{S}$-$\mathrm{DC}$ cuts when $m$ nodes remain is denoted as $(n, m)$, and the capacity set of that is denoted as $c(n, m)$, $k \leq m \leq n - 1$.
In $(n, m)$, each cut is determined by a combination of edges $\mathrm{N_u} \rightarrow \mathrm{N_u}$ and $\mathrm{N_u} \rightarrow \mathrm{N_w}$ with capacity $\alpha$ and $\beta$, respectively.
For calculating the cut (capacity), we do not explicitly consider in the transmissions sent from the nodes that survive until a collector comes to reconstruct data.
In other words, only transmissions sent from the nodes which do not survive are beneficial for data protection, because they are needed to deliver the information (degrees of freedom) to the remaining nodes.
As explained, the symmetry in our problem and the fact that we account for the cost of transmissions from all nodes at each repair phase, allows us to focus on a specific sequence of lost nodes without compromising the accuracy of the result.
Based on this, we formulate $c(n, m)$ as

\footnotesize
\begin{equation}
\begin{aligned}
   c \left(n, m \right)
    = & \{j_n \alpha_n + j_{n - 1}\alpha_{n - 1} + \cdots + j_m \alpha_m + \\
    & l_n\beta_n + l_{n - 1}\beta_{n - 1} + \cdots + l_{m + 1} \beta_{m + 1} \vert \\
    & j_{n}, \ldots, j_m \in \mathbb{N} \wedge j_{n}, \ldots, j_m \leq m, \\
    & l_n, \ldots, l_{m + 1} \in \mathbb{N} \} ~,
\label{nm}
\end{aligned}
\end{equation}
\normalsize
where $\mathbb{N}$ is the set of all natural numbers.

While $m$ nodes are remaining and connected to $\mathrm{DC}$, there must be $m$ edges $\mathrm{N_u} \rightarrow \mathrm{N_u}$ in each $\mathrm{S}$-$\mathrm{DC}$ cut, which means the total number of $\alpha$ in each cut capacity is $m$, namely,
\vspace{-0.1cm}
\begin{equation}
\sum_{p = m}^n j_p = m ~.
\label{jp}
\vspace{-0.1cm}
\end{equation}
As stated above, $(q - m)$ nodes send beneficial transmissions in a repair stage involving $q$ nodes $(m + 1 \leq q \leq n)$. Regarding this stage, the number of beneficial transmissions in a cut is also dependent on the number of edges $\mathrm{N_u} \rightarrow \mathrm{N_u}$ that have been counted in the cut, which is $\sum_{p = q}^n j_p$. Specifically, the beneficial transmissions received by $\sum_{p = q}^n j_p$ nodes are in the cut.  
Thus, the number of edges $\mathrm{N_u} \rightarrow \mathrm{N_w}$ in a cut is
\vspace{-0.1cm}
\begin{equation}
l_q = (q - m)\sum_{p = q}^n j_p ~.
\label{lq}
\vspace{-0.1cm}
\end{equation}
By \eqref{nm}, \eqref{jp} and \eqref{lq}, we acquire the capacity set of every stage $c(n, m)$ for a given pair of $n$ and $k$.
\begin{figure*}[ht]
    \centering
    \begin{subfigure}{0.23\textwidth}
        \includegraphics[width=\linewidth, height = 2.53cm]{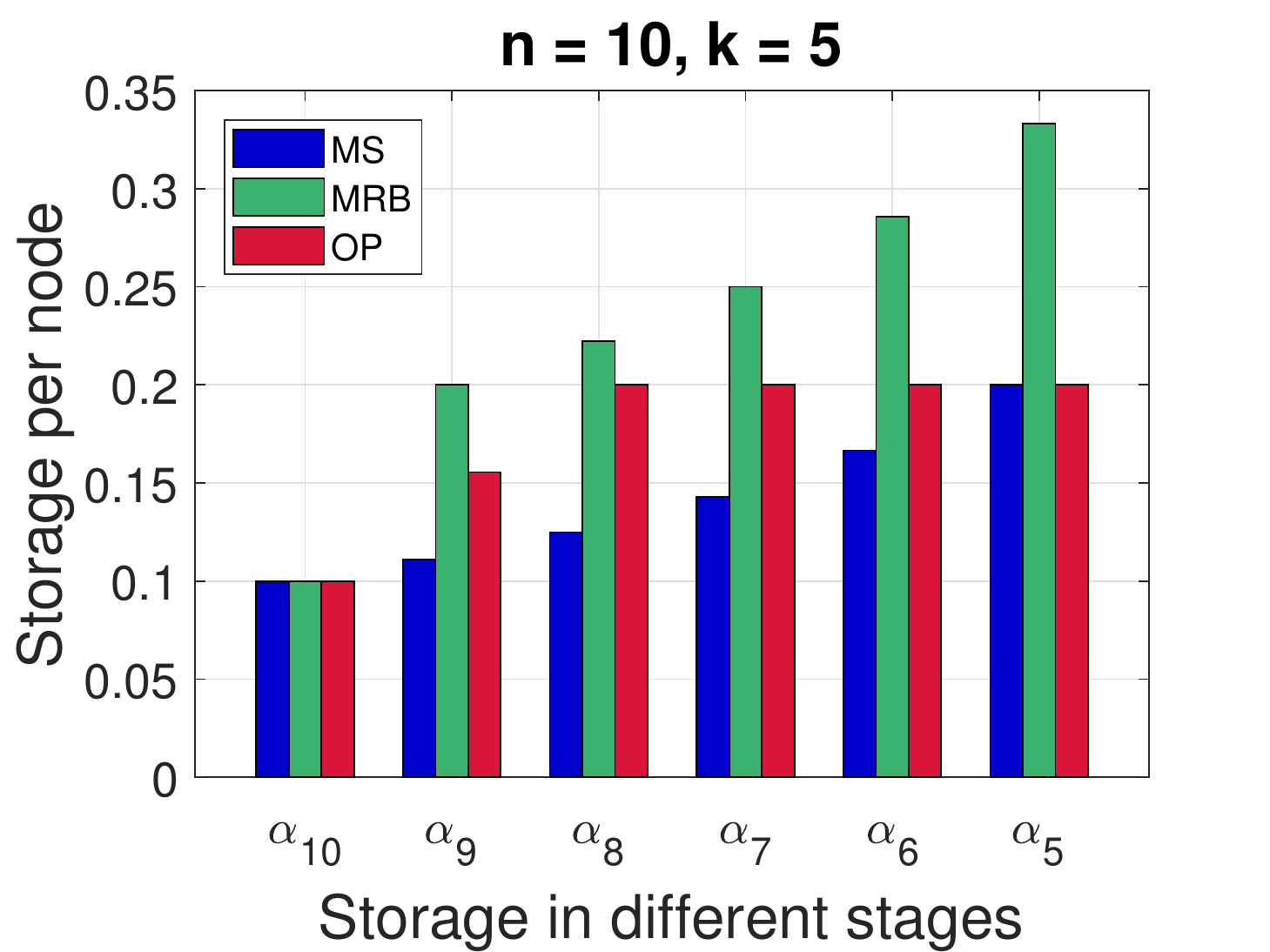}
        \caption{Storage in different stages for $n = 10,k = 5$.}
        \label{fig3a}
    \end{subfigure}
    \begin{subfigure}{0.23\textwidth}
        \includegraphics[width=\linewidth, height = 2.53cm]{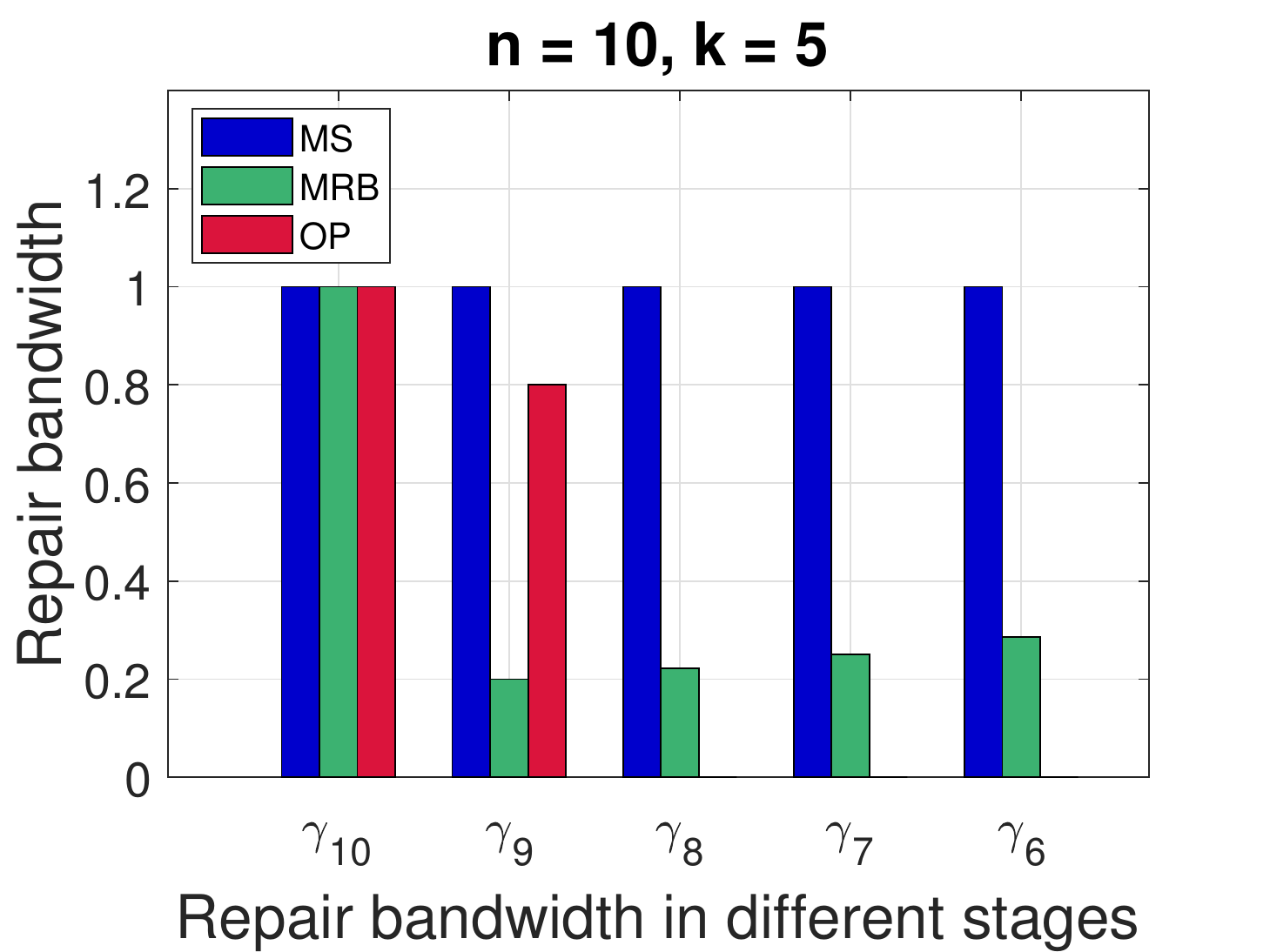}
        \caption{Repair bandwidth in each stage for $n = 10,k = 5$.}
        \label{fig3b}
    \end{subfigure}
    \begin{subfigure}{0.23\textwidth}
        \includegraphics[width=\linewidth, height = 2.53cm]{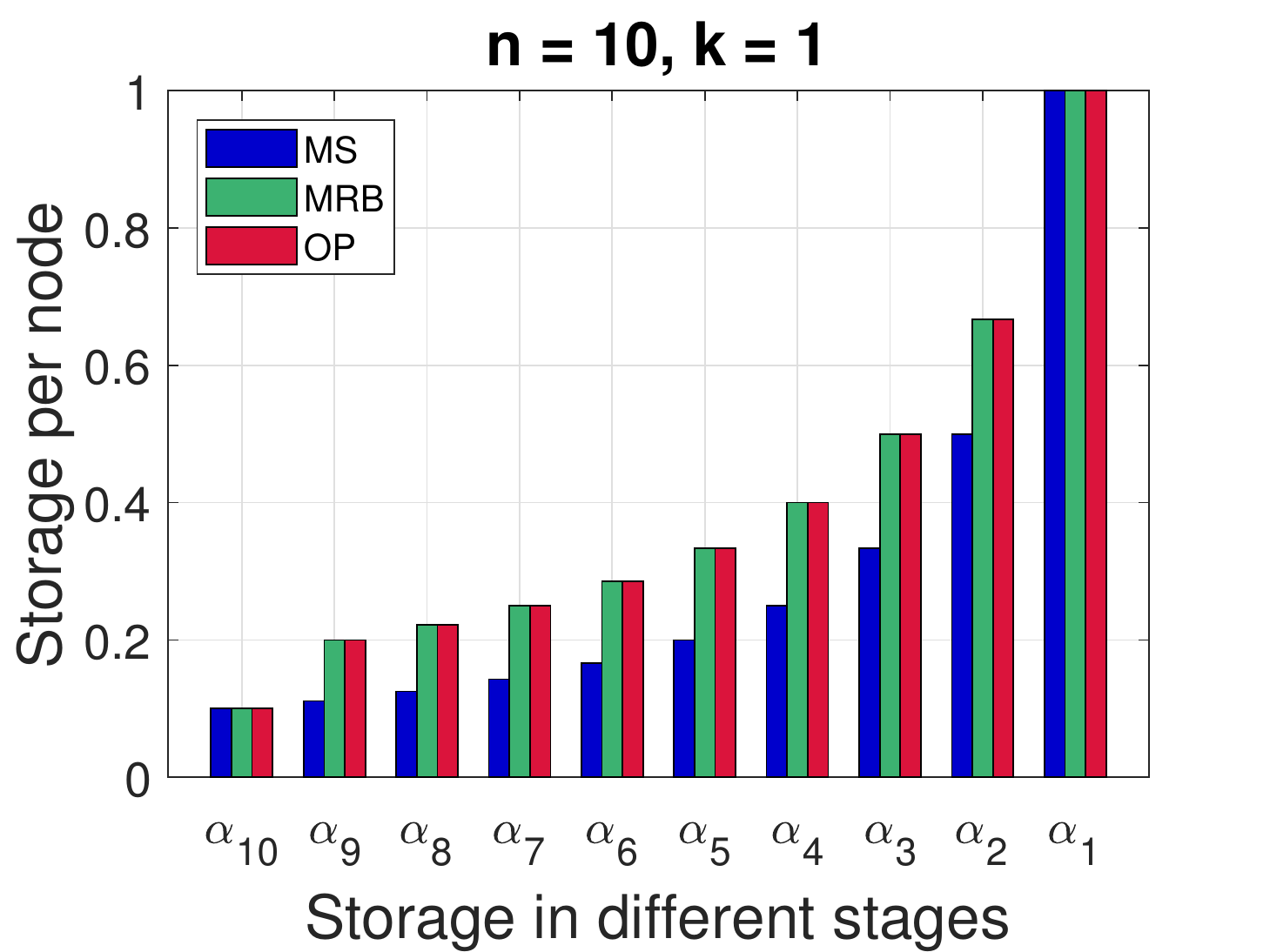}
        \caption{Storage in different stages for $n = 10,k = 1$.}
        \label{fig3c}
    \end{subfigure}
    \begin{subfigure}{0.23\textwidth}
        \includegraphics[width=\linewidth, height = 2.53cm]{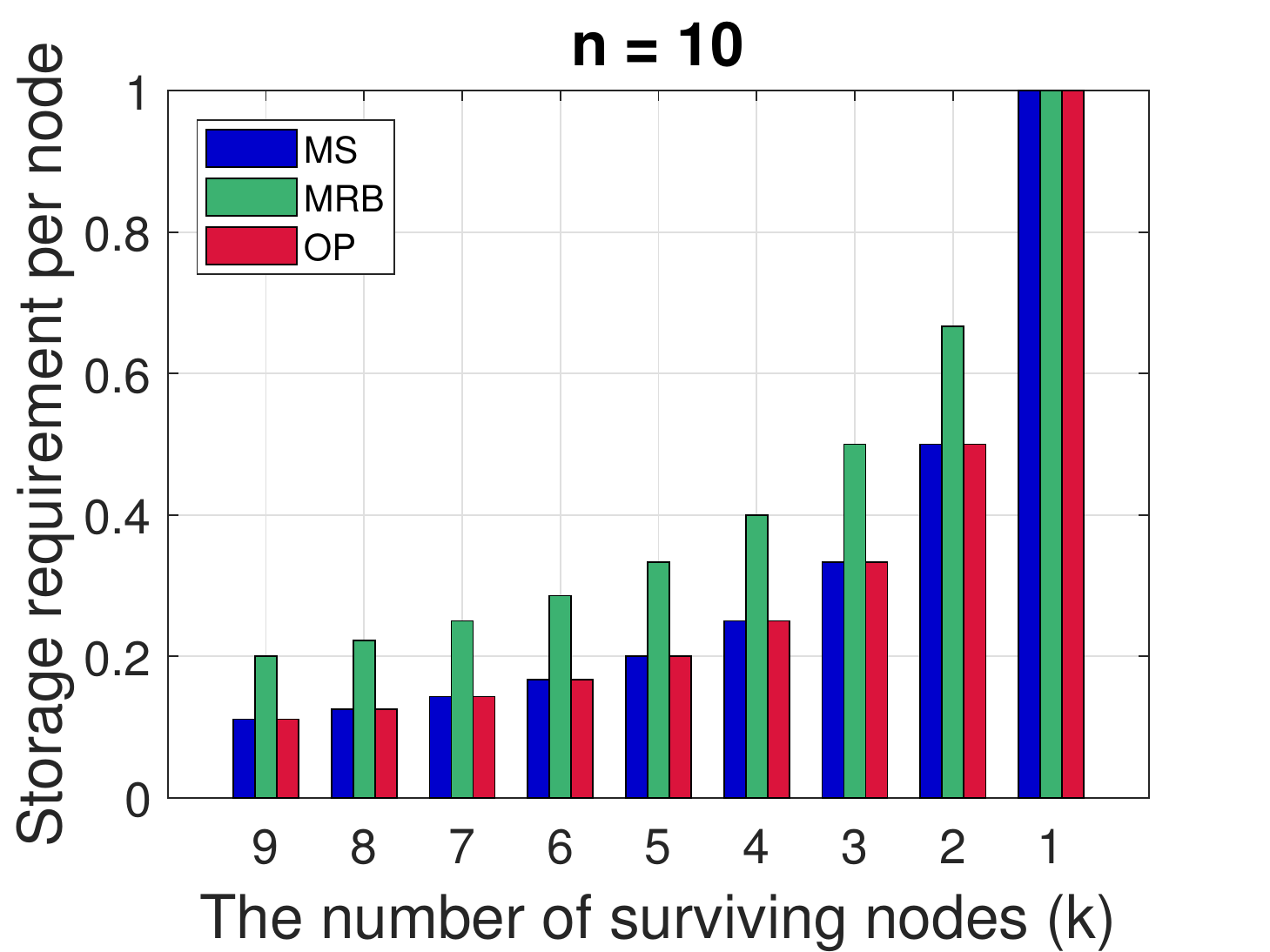}
        \caption{Storage requirement for ~different $k$, and $n = 10$. }
        \label{fig3d}
    \end{subfigure}
    \caption{Detailed metrics of different strategies \vspace{-0.4cm}}\label{Fig3}
\vspace{-0.2cm}
\end{figure*}

Minimizing the protection bandwidth can generate multiple feasible solutions for the storage requirements. Accordingly, we modify slightly the cost function to select the min protection bandwidth that requires the least amount of storage.
For this reason, we use a cost function the protection bandwidth $\delta$ plus a small term concerning storage overhead $\varepsilon\cdot\sigma$ as the objective function, where $\varepsilon > 0$ is a very small constant (e.g., $10^{-6}$). In this way, the optimal protection strategy with minimum storage overhead is selected. Therefore, we formulate the optimization problem as follows,

\footnotesize
\begin{equation}
\begin{aligned}
\quad & \quad \quad \quad \text{min} \quad \delta(n, k) + \varepsilon\cdot\sigma(n, k) \\
&\text{s.t.}
\begin{cases}
\bigcup\limits_{m = k}^{n - 1} c(n, m) \geq M \\
\alpha_{m + 1} + m \beta_{m + 1} \geq \alpha_m & m = k,\ldots, n - 1 \\
0 \leq \beta_{s} \leq \alpha_{s} & s = k + 1,\ldots, n \\
\alpha_n = \frac{M}{n} ~,
\end{cases}
\end{aligned}
\label{op_nk}
\end{equation}
\normalsize
while maintaining a resulting $\delta(n, k)$ that is the same as solving the problem for min $\delta(n, k)$. Since \eqref{op_nk} is a formulation of \eqref{op_64}, we do not repeat the analysis in the example.
By solving this LP problem, we determine the minimum protection bandwidth $\delta(n, k)_{min}$, and the optimal protection strategy $(\beta_{k + 1}, \ldots, \beta_n)$ with least storage overhead.

The computational complexity of an LP depends on the number of variables and the input size. As to our LP problem, the number of variables is $2(n - k)$, and the size of cut capacity constraints is $\left| \bigcup\limits_{m = k}^{n - 1} c(n, m) \right| = \sum_{t = 1}^{n - k} \binom{n}{t}$, which increases dramatically and makes the input size enlarge with $n$ increasing, especially when the difference of $n$ and $k$ is large. So, moderate values for them are expected from a practical perspective in IoT and monitoring problems. Approximated solutions will be considered in future works for large values.
\vspace{-0.15cm}

\vspace{-0.15cm}
\section{Evaluation}

In this section, we compare the optimal strategy with two suboptimal strategies inspired by the work in~\cite{guerrero2016network}.

$Optimal \ Protection \ strategy (OP)$: OP is the solution the optimization problem in~\eqref{op_nk}.

$Minimum \ Storage \ strategy (MS)$: MS is a protection strategy inspired by the MSR strategy in~\cite{guerrero2016network}, which stores minimum data after each repair stage.

$Minimum \ Repair \ Bandwidth \ strategy (MRB)$: MRB is inspired by the MBR repair strategy in~\cite{guerrero2016network}, which transitions between MBR storage points in each node given by the results in~\cite{dimakis2010network} and the number of remaining nodes. This scheme focuses on optimizing repair bandwidth in each stage (repair triggered after every loss). However, it disregards the cost of the early protection phase and provides a myopic solution for the protection process.

Although the system model in this letter is different from those in~\cite{dimakis2010network, guerrero2016network}, the operating points $(\alpha_{MSR}, \gamma_{MSR})$ and $(\alpha_{MBR}, \gamma_{MBR})$ described in the previous results can still be achieved by MS and MRB, respectively, which makes the strategies in this letter comparable to those in~\cite{guerrero2016network}. We omit the proof as it is beyond the scope of this work.

In plots, the protection bandwidth and storage are normalized with the size of original data $M$.
Fig.~\ref{fig2a} shows that the protection bandwidth $\delta$ of MS increases rapidly and linearly when the number of surviving nodes $k$ decreases. This happens due to the fact that the bandwidth in each repair stage is large, specifically, $M$ bits in order to guarantee data recoverability.
Compared with MRB, OP has a significant reduction in terms of protection bandwidth when $k$ is between $8$ and $3$. The reduction is up to $17$\% of $M$, and over $12$\% of $M$ on average. In particular, adopting OP will save considerable network resources when $M$ is larger.
For the case of $k = n - 1 = 9$, the storage before repair is $M / n$, i.e., the minimum storage irrespective of the value of $n$, and the protection bandwidth is the same as the repair bandwidth, i.e., $M$.
Note that MRB has the same performance as OP on $\delta$ if $k = 2$ and $k = 1$, which means that MRB is optimal under those conditions. Fig.~\ref{fig2b} shows $\delta(n, k)_{min}$ for different $n$. Considering a certain $n - k$, a greater $n$ leads to smaller $\delta(n, k)_{min}$, due to the fact that a greater $n$ indicates greater $k$. As more nodes participate in a repair stage, the number of bits per transmission decreases.

Let us understand the process of repair. Fig.~\ref{fig3a} shows that for $n = 10$ and $k = 5$ the storage of MS raises gradually, while that of OP raises faster in the first two stages. Fig.~\ref{fig3b}, the corresponding repair bandwidth of Fig.~\ref{fig3a}, shows that OP transmits more in the first two repair stages, but, it transmits much less in future repairs as it stores more in each node in that first stage. Fig.~\ref{fig3c} shows that MRB is optimal (matches OP) in terms of $\delta$ and $\alpha$ when $k = 1$.
Finally, Fig.~\ref{fig3d} shows that the storage requirements of OP are as low as those of MS, and are much less than those of MRB. The reduction is up to 16.6 \% of $M$ and 11.4 \% of $M$ on average for $2 \leq k \leq 9$.
\vspace{-0.1cm}

\section{Conclusion}
\vspace{-0.1cm}

This letter formulated the information flow approach to solve the problem of efficient repair of losses without newcomer nodes, which is an important problem in IoT and monitoring applications. We showed that the optimal strategy to reduce communication bandwidth outperforms a prior scheme from~\cite{guerrero2016network} not only in bandwidth use, but also in storage use. Although the optimal solution provided is an LP that can be solved for moderate $n$ and $k$ values in the network, the number of min-cut constraints might hinder its use for $n > 100$ nodes. Although in practice this can be handled by establishing clusters of nodes of reasonable size, our future work will study relaxations to the min-cut constraints in order to solve the problem for large $n$. Future work will consider practical implementations as well as extending our formulation to the case of a lead repair node used in~\cite{guerrero2016network,zhao2017reliable}.
\vspace{-0.15cm}

\ifCLASSOPTIONcaptionsoff
  \newpage
\fi



\bibliographystyle{IEEEtran}
\bibliography{IEEEabrv, ReliableIoTStorage}

\end{document}